\documentstyle[twocolumn,aps,prl]{revtex} 
\begin{document} 
\def\irrep{irreducible representation}
\def\irreps{irreducible representations}
\def\t{\times}
\def\<{\langle}
\def\>{\rangle}
\title{ Comment on "Evidence for the Existence of Supersymmetry
in Atomic Nuclei"}

\author{ B G Wybourne\cite{e}\cite{w} 
}
\address{Instytut Fizyki, UMK,
ul. Grudzi\c{a}dzka 5/7,87-100 Toru\'n, Poland}
\maketitle 


Metz {\it et al}\cite{1.} have reported evidence for the existence of
supersymmetry in atomic nuclei by consideration of the nuclear
quartet $^{194}Pt$, $^{195}Pt$, $^{195}Au$, $^{196}Au$. Their
states are derived from a selection of the states that arise in
the decomposition of the 705942-dimensional \irrep\ $[5]\t[2]$
of the supersymmetry product group $U(6/12)\t U(6/4)$ under
restriction via the group chain\cite{2.}

\begin{equation}
\begin{tabular}{l}
$U_\nu(6/12)\t U_\pi(6/4)$\\
$\sqcup$\\
$U_\nu^F(12)\t U_\nu^B(6)\t U_\pi^B(6)\t U_\pi^F(4)$\\
$\sqcup$\\
$SU_\nu^F(2)\t U_\nu^F(6)\t U_{\nu+\pi}^B(6)\t SU_\pi^F(4)$\\
$\sqcup$\\
$SU_\nu^F(2)\t U_{\nu+\pi}^{B+F}(6)\t SU_\pi^F(4)$\\
$\sqcup$\\
$SU_\nu^F(2)\t O_{\nu+\pi}(6)\t SU_\pi^F(4)$\\
$\sqcup$\\
$SU_\nu^F(2)\t SO_{\nu+\pi}(6)$\\
$\sqcup$\\
$SU_\nu^F(2)\t SO_{\nu+\pi}(5)$\\
$\sqcup$\\
$SU_\nu^F(2)\t SO_{\nu+\pi}(3)$\\
$\sqcup$\\
$Spin(3)$\\
\end{tabular}
\end{equation}

They introduce a model Hamiltonian involving a linear combination of
the second-order Casimir operators of the groups appearing in the
group chain leading to the energy eigenvalue expression for the case
in mind

\begin{equation}
\begin{tabular}{l}
$E=A[N_1(N_1+5)+N_2(N_2+3)]$\\
$+B[\Sigma_1(\Sigma_1+4)+\Sigma_2(\Sigma_2+2)]$\\
$+B^\prime[\sigma_1(\sigma_1+4)+\sigma_2(\sigma_2+2)+\sigma_3^2]$\\
$+C[\tau_1(\tau_1+3)+\tau_2(\tau_2+1)]+DL(L+1)+EJ(J+1)$\\
\end{tabular}
\end{equation}

with $A,\ B,\ B^\prime,\ C,\ D,$ and $E$ free parameters and the
$[N_1,N_2],\<\Sigma_1,\Sigma_2\>,\<\sigma_1,\sigma_2,\sigma_3\>,
(\tau_1,\tau_2),L,J$ quantum numbers correlated to the \irreps\
of $U(6),O(6),SO(6),SO(5),SO(3)$ and spin(3) respectively.

Metz {\it et al} appear to have overlooked a subtle point that
must invalidate much of their analysis, and indeed earlier analyses\cite{2.}\cite{3.}, - the expression in (2)cannot distinguish 
\irreps\ occurring in the decomposition
along the assumed group chain (1) with a multiplicity greater than
unity. This is no problem for the states they have based upon the
$[7,0]$ and $[6,0]$ \irreps\ of $U_{\nu+\pi}^{B+F}(6)$ but it is
a serious problem for those states based upon the \irreps\
$[6,1]$ and $[5,1]$ where states of multiplicity 2 occur.

The problem can be seen by considering the quadruple group product
$SU_\nu^F(2)\t U_\nu^F(6)\t U_{\nu+\pi}^B(6)\t SU_\pi^F(4)$. 
To contract to the group $U_{\nu+\pi}^{B+F}(6)$ we must form the
Kronecker product of the \irreps\ of the two $U(6)$ groups.
Consider the two quadruple product \irreps\

\begin{equation}[1]\t[1]\t[6]\t[0],\quad [1]\t[1]\t[5,1]\t[0]\end{equation}

In contracting the two $U(6)$ \irreps\ we need the two Kronecker products

\begin{equation}
[1]\t[6] = [7] + [6,1],  [1]\t[5,1] = [61] + [5,2] + [5,1,1]
\end{equation}

This shows unambiguously that there arise two distinct $[6,1]$
\irreps\ which in $U_{\nu+\pi}^{B+F}(6)\t SO_\pi^F(6)\t SU_\nu^F(2)$
are ambiguously given as
$$[6,1]\t\<0\>\t[1]$$
which cannot be distinguished by the Metz {\it et al} Hamiltonian. In
an exactly similar manner one can show that the $[5,1]$ is also
associated with a multiplicity of 2. To distinguish the states with
multiplicity greater than unity requires modification of (2), possibly
by the inclusion of a term involving the product of the Casimir
operators for the two $U(6)$ groups. Without separating the terms one
has no assurance that one of the \irreps\ could not be driven lower and
that the degeneracies are properly described.  

\subsection*{Acknowledgement} 

This work was been partially supported
by Polish KBN Grant 2 P03B 076 13 and the Max Planck Institut f\"ur Astrophysik (Garching).

\end{document}